\documentclass[runningheads]{llncs}
\usepackage[T1]{fontenc}
\usepackage{booktabs}
\usepackage{graphicx,verbatim}
\usepackage{bm}
\usepackage{marvosym}
\usepackage{multirow}
\usepackage{amsfonts,amssymb}
\usepackage{amsmath}
\usepackage{hyperref}
\hypersetup{colorlinks=true, linkcolor=blue, anchorcolor=blue, citecolor=blue, urlcolor=blue}
\usepackage[symbol]{footmisc}
\usepackage{graphicx}
\begin{document}
\title{Ophora: A Large-Scale Data-Driven Text-Guided Ophthalmic Surgical Video Generation Model}
\titlerunning{Ophora}

\authorrunning{W. Li et al.}

\newcommand*\samethanks[1][\value{footnote}]{\footnotemark[#1]}
\author{Wei Li \inst{1,2} \textsuperscript{*} \and
Ming Hu \inst{1,3} \textsuperscript{*}  \and
Guoan Wang\inst{4}  \and
Lihao Liu\inst{1}  \and
Kaijing Zhou\inst{5}  \and
Junzhi Ning\inst{1,7}  \and
Xin Guo\inst{6}   \and
Zongyuan Ge\inst{3}   \and
Lixu Gu\inst{2}   \and
Junjun He\inst{1,8}  \textsuperscript{$\dagger$} }

\institute{Shanghai Artificial Intelligence Laboratory, China \\ \email{hejunjun@pjlab.org.cn}  \and
Shanghai Jiao Tong University, China \\ \email{liwei2022@sjtu.edu.cn} \and
Monash University, Australia \and East China Normal University, China \and Eye Hospital, Wenzhou Medical University, China \and Shanghai Academy of Artificial Intelligence for Science, China  \and
Imperial College London, UK \and Shanghai Innovation Institute, China
}

\maketitle              % typeset the header of the contribution

\footnotetext[1]{\quad  Equal contribution}
\footnotetext[2]{\quad  Corresponding Author}

\begin{abstract}
In ophthalmic surgery, developing an AI system capable of interpreting surgical videos and predicting subsequent operations requires numerous ophthalmic surgical videos with high-quality annotations, which are difficult to collect due to privacy concerns and labor consumption.
Text-guided video generation (T2V) emerges as a promising solution to overcome this issue by generating ophthalmic surgical videos based on surgeon instructions.
In this paper, we present Ophora, a pioneering model that can generate ophthalmic surgical videos following natural language instructions. 
To construct Ophora, we first propose a Comprehensive Data Curation pipeline to convert narrative ophthalmic surgical videos into a large-scale, high-quality dataset comprising over 160K video-instruction pairs, Ophora-160K. 
Then, we propose a Progressive Video-Instruction Tuning scheme to transfer rich spatial-temporal knowledge from a T2V model pre-trained on natural video-text datasets for privacy-preserved ophthalmic surgical video generation based on Ophora-160K.
Experiments on video quality evaluation via quantitative analysis and ophthalmologist feedback demonstrate that Ophora can generate realistic and reliable ophthalmic surgical videos based on surgeon instructions. We also validate the capability of Ophora for empowering downstream tasks of ophthalmic surgical workflow understanding. 
Code is available at \url{https://github.com/uni-medical/Ophora}.

% Authors must provide keywords and are not allowed to remove this Keyword section.

\end{abstract}

\section{Introduction}
In ophthalmic surgery, surgical scenes are often recorded as videos \cite{THIA2019570videorecord}. 
AI systems capable of interpreting ophthalmic surgical videos and predicting subsequent operations hold the potential to improve procedural performance and reduce postoperative infections, especially when integrated with surgical robotics \cite{aiinsurg,cepolina2024review,cheng2023deep,he2021robotoph}.
However, developing such systems requires numerous surgical videos with high-quality annotations, which are difficult to acquire due to privacy concerns \cite{price2019privacy} and labor consumption \cite{laborconsump}.
To overcome this issue, generating surgical videos based on surgeon requirements emerges as a promising solution.

Text-guided video generation (T2V) \cite{t2voriginal,yang2024cogvideox} provides a feasible method for generating ophthalmic surgical videos from natural language instructions. 
Current T2V work relies on surgical phase labels \cite{cho2024surgentextguideddiffusionmodel,Laparoscopicgen}, which typically lack detailed descriptions for accurate surgical video generation. \textit{As a result, these models struggle to capture fine-grained actions and intricate interactions between instruments and anatomical structures} \cite{miradata}. 
Additionally, existing T2V research has also explored utilizing pre-trained Text-to-Image models with temporal mixing layers \cite{endora,bora}, which capture spatial and temporal correspondences separately. \textit{However, this approach fails to holistically consider the complex spatial-temporal relationships inherent in surgical videos, leading to frame inconsistency} \cite{yang2024cogvideox}. 

To address the first challenge, our primary motivation is to curate a large-scale, high-quality collection of surgical video-text pairs. Inspired by the properties of open-source narrative surgical videos \cite{hu2024ophcliphierarchicalretrievalaugmentedlearning,hecvl}, which provide detailed descriptions, including surgical phases, instruments, and medications from professional surgeons, we aim to collect similar videos from the internet and convert them into video-text pairs.
To address the second challenge, we focus on directly modeling spatial-temporal correspondence for improved frame consistency. Inspired by the success of transfer learning \cite{transferlearning} and acknowledging the significant scale gap between natural and surgical videos \cite{hecvl}, we propose to transfer the spatial-temporal knowledge from a T2V model pre-trained on large-scale natural video-text datasets to guide the generation of surgical videos. 
During the transfer learning process, we also address the challenge of privacy preservation by ensuring that the generated videos exclude sensitive visual content, such as subtitles and watermarks, that are unrelated to the surgical process.

In this paper, we propose a novel text-guided ophthalmic surgical video generation model, named \emph{Ophora}, that can generate realistic and reliable ophthalmic videos following natural language instructions.
To achieve this, we first propose a \emph{comprehensive data curation} pipeline to construct a large-scale, high-quality video-instruction dataset, \emph{Ophora-160K}, which includes over 160K video clips paired with generation instructions. The pipeline involves eliminating irrelevant narrative information and filtering clips with extreme dynamics. 
Next, we propose a \emph{progressive video-instruction tuning} approach to develop Ophora from a T2V model that learns spatial-temporal knowledge from natural videos. 
Specifically, we conduct transfer pre-training on this T2V model to transfer its knowledge for ophthalmic video generation using Ophora-160K. Then, we fine-tune the model on clips without sensitive information for preserving privacy. 
We conduct experiments to evaluate Ophora by assessing quality of synthesized videos and potential for downstream ophthalmic surgical workflow understanding tasks \cite{hu2024ophnetlargescalevideobenchmark}.

% Our contributions are three-fold:

% \begin{itemize}
%     \item We propose a comprehensive data curation pipeline for creating a high-quality, large-scale video-text dataset, \emph{Ophora-160K}, which contains over 160K video clips with generation instructions. The pipeline focuses on removing irrelevant narrative information and filtering clips with extreme dynamics.
    
%     \item We introduce \emph{Ophora}, a Progressive Video-Instruction Tuning approach that can generate realistic 6-second ophthalmic videos at a frame rate of 8 FPS based on text instructions. Our method involves transfer pre-training on a T2V model that directly learns spatial-temporal correspondence from natural videos, followed by fine-tuning on privacy-preserving clips devoid of irrelevant visual content. \textit{To the best of our knowledge, we are the first to achieve this approach.}
    
%     \item We conduct extensive experiments to evaluate Ophora's capabilities, assessing the quality of synthesized videos from both subjective and objective perspectives, and demonstrating its effectiveness in improving the performance of existing AI models on downstream tasks such as ophthalmic surgical phase recognition \cite{hu2024ophnetlargescalevideobenchmark}.
% \end{itemize}

% 补一个整体方法示意图
% 把downstream 单独画出来，不放在quality validation里，修改一下其中的一些单词

\section{Method}
%先简要介绍一下我们的方法的思路
% 今天要把Method部分重新写完一下
% 1. 构建了一个数据集；2. 训练了一个模型
%Since there is no readily available large-scale, high-quality video-text dataset for ophthalmic surgical video generation, we first construct such a dataset from narrative videos (Sec. \ref{sec:data-curate}). 

\begin{figure}[!t]
    \includegraphics[width=\textwidth]{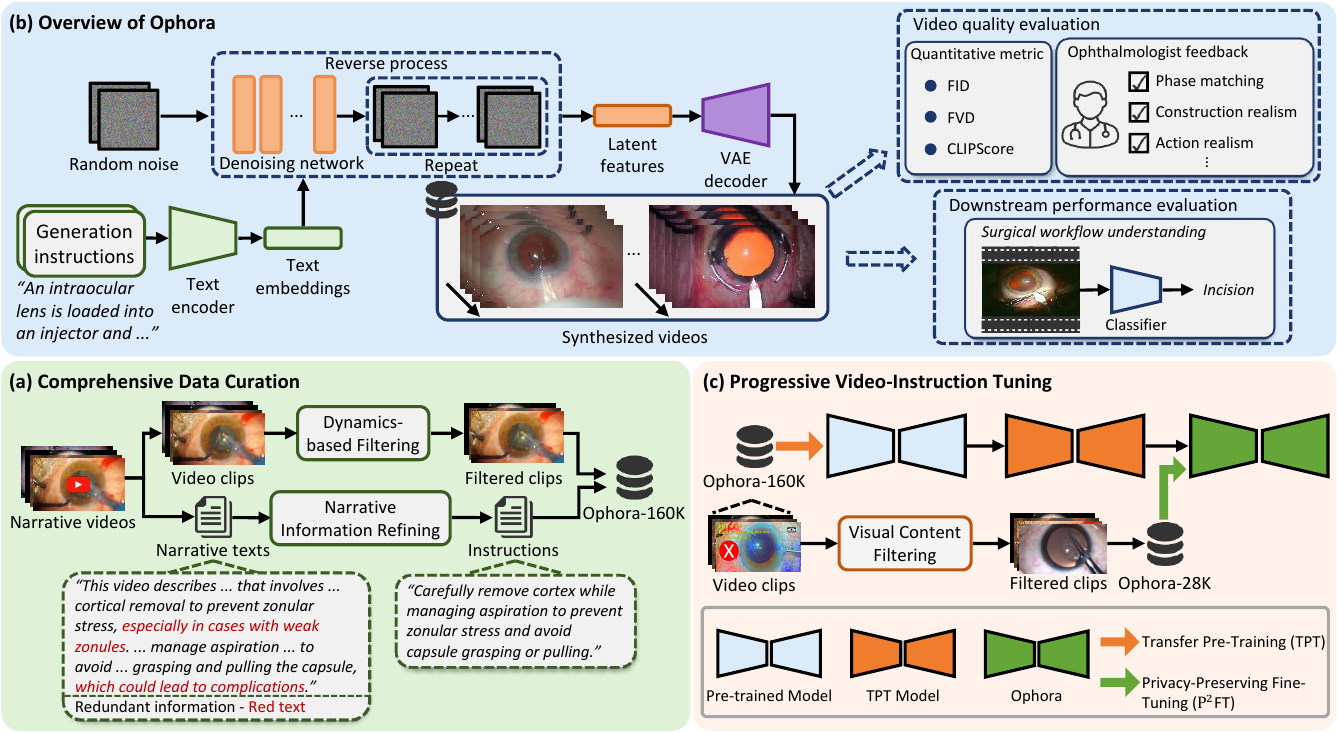}
    \caption{Illustration of proposed Ophora that can generate ophthalmic surgical videos from instructions. Specifically, we propose a Comprehensive Data Curation pipeline to construct a large-scale, high-quality video-instruction dataset, Ophora-160K, from narrative videos (Sec. \ref{sec:data-curate}). We introduce a T2V model pre-trained on natural video-text pairs (Sec. \ref{sec:model}) and leverage Progressive Video-Instruction Tuning to transfer spatial-temporal knowledge from the pre-trained model for ophthalmic video generation while preserving privacy using Ophora-160K (Sec. \ref{sec:tuning}). We evaluate the capabilities of Ophora by assessing synthesized video quality and downstream performance.}
    \label{fig:overall}
\end{figure}

In this section, we first construct a large-scale, high-quality video-text dataset from narrative videos (Sec. \ref{sec:data-curate}).
Then, we introduce a text-guided video generation (T2V) model pre-trained on natural video-text datasets as the generator backbone (Sec. \ref{sec:model}) and apply the proposed Progressive Video-Instruction Tuning to transfer the spatial-temporal knowledge for ophthalmic video generation while preserving privacy (Sec. \ref{sec:tuning}). 
Fig. \ref{fig:overall} illustrates the overall framework.

\subsection{Comprehensive Data Curation}

\label{sec:data-curate}
% 写一下数据的处理、筛选流程等
OphVL \cite{hu2024ophcliphierarchicalretrievalaugmentedlearning} collects large-scale ophthalmic surgical narrative videos from the Internet and converts them into video clip-caption pairs.
However, these captions contain redundant narrative information, e.g., educational content, that indirectly corresponds to the visual content and affects the instruction-following capabilities of a T2V model \cite{miradata}.
Besides, these videos exhibit variable temporal dynamics, e.g., intense camera shaking or nearly static content due to differences in surgeon experience, leading to temporal incoherence \cite{dynamicsbench}.
Therefore, we propose a comprehensive data curation pipeline to refine OphVL from textual and visual sides for video generation, as shown in Fig. \ref{fig:overall}(a).

\noindent
\textbf{Narrative Information Refining.}
%However, OphVL retains the entire original narrative information, including details indirectly unrelated to the surgical activities, which must be effectively removed with the remaining content organized into coherent natural language.
%说明使用LLM过滤
Given the remarkable text understanding capabilities of existing large language models (LLMs), we employ a powerful open-source LLM, Qwen2.5-72B \cite{qwen25}, to remove the irrelevant information from the captions and transform them into generation instructions.
We manually select 10 captions and carefully annotate the redundant information within them as examples. Then, we leverage these examples to instruct the LLM to refine the remaining captions to enhance the accuracy and stability of the LLM's outputs.

%visual 内容的过滤
\noindent
\textbf{Dynamics-based Filtering.}
%这一块可以考虑用数学公式重述，描写具体是基于什么变量的阈值来挑选keyframes
%On the other hand, during the training process, videos with large dynamics increase the difficulty of spatial-temporal information extraction, while videos with small dynamics (nearly still) containing sparse temporal information can lead to synthesizing videos with barely noticeable activities.
%考虑用含公式的式子重写一下
We design a simple rule to filter clips with poor temporal dynamics quality. We utilize PySceneDetect toolkit to extract keyframes from a clip at timestamps where significant visual changes occur. 
Clips with an excessive or insufficient number of keyframes are filtered based on predefined thresholds. We empirically set the upper and lower thresholds to 100 and 2.

\noindent
\textbf{Ophora-160K.}
We further filter out low-resolution clips (below 720$\times$480) and ultimately construct a large-scale video-instruction dataset, Ophora-160K, comprising approximately 160K clip-level video-instruction pairs.

\subsection{Overview of Ophora}
\label{sec:model}
% 介绍text to video generation model
% 添加一些数学符号表述，对模型的输入和输出
The backbone of Ophora is based on CogVideoX-2b \cite{yang2024cogvideox}, a latent diffusion model with strong capability of generating natural videos from text, shown in Fig. \ref{fig:overall}(b). 
It consists of a 3D Variational Autoencoder (VAE), a T5 text encoder \cite{t5encoder}, and a transformer-based denoising network. 
Given a video $x\in \mathbb{R}^{F\times H\times W\times C}$, where $F, H, W,C$ denote the number of video frames, height, width, and channel number, respectively, the video is compressed into latent space using the VAE encoder.
The latent representations are then patchified and flattened into $D$-dimensional vision embeddings $z^v$ of length $\frac{F}{q}\cdot \frac{H}{p} \cdot \frac{W}{p}$, where $q, p$ denote the temporal and spatial compression rates.
Meanwhile, an input text is encoded into text embeddings $z^\phi$ via the T5 encoder. 
Then, random Gaussian noise $\epsilon\sim\mathcal{N}(\mathbf{0},\mathbf{I})$ is added to $z^v$, i.e., $z^v_t = \sqrt{\bar{\alpha}_t} z^v + \sqrt{1-\bar{\alpha}_t} \epsilon$ with a noise level $\alpha_t$ at a random timestep $t\in[1:T]$, where $T$ denotes the total diffusion steps, during the diffusion process \cite{ddpm}. Subsequently, $z^v_t$ and $z^\phi$ are concatenated into a sequence and fed into the denoising network $\epsilon_\theta$ to predict the Gaussian noise $\epsilon$. 
The model is trained through the optimization objective of diffusion models:
\begin{equation}
    L_{\text{diff}} = {\mathbb{E}}_{t,\epsilon_t,(z^v,z^\phi)\sim\mathcal{D}}\Big [ \Vert {\epsilon} - {\epsilon}_\theta ([z^v_t,z^\phi], t) \Vert^2_2 \Big ],
    \label{eq:T2V}
\end{equation}
where $\mathcal{D}$ denotes the training dataset and $[,]$ is concatenation operation.
After training, we sample random noise and use $\epsilon_\theta$ in reverse process to iteratively generate latent vision embeddings from the noise given a text input. Subsequently, the vision embeddings are reconstructed to a video via the VAE decoder.

\subsection{Progressive Video-Instruction Tuning}
\label{sec:tuning}
% 写一下每个阶段训练的一些内容
A T2V model pre-trained on large-scale natural videos provides reliable prior spatial-temporal knowledge for generating ophthalmic surgical videos.
We aim to transfer such knowledge for privacy-preserved ophthalmic video generation.
However, most clips in Ophora-160K contain sensitive visual information, such as subtitles or watermarks, prompting us to adopt a two-stage training approach, where we first introduce ophthalmic knowledge into the model and then guide it to generate videos without such sensitive information, as shown in Fig. \ref{fig:overall}(c).
%In the first stage, we continue to pretrain the T2V model using all Ophora-160K video-instruction pairs for comprehensive knowledge integration. 
%In the second stage, we fine-tune the stage-one model on the pairs with clean video clips for privacy preserving and reliable video generation. Clean clips exclude subtitles, watermarks, microscope windows or parameters during the recording, and other visual information not directly related to surgical activities [xx].

\noindent
\textbf{Transfer Pre-Training.}
We utilize the entire Ophora-160K dataset to conduct continual pre-training on the backbone, focusing solely on training the denoising network based on the Eq. \ref{eq:T2V} while keeping T5 and the VAE frozen. 
We uniformly divide $[1:T]$ into $N$ sub-intervals when training the model on $N$ GPUs and sample $t$ from each sub-interval on each GPU to improve training efficiency.
%理由, 说明为什么不更新VAE, 感觉这里不能说实验证明，没有空间了
%We conduct preliminary experiments and verify that the VAE achieves satisfactory reconstruction performance on ophthalmic videos (Sec. xx). Therefore, we choose not to train the VAE to improve training efficiency.

\noindent
\textbf{Privacy-Preserving Fine-tuning.} 
%Our dataset contains clips with content indirectly related to surgical activities as mentioned above, and therefore needs to be further filtered.
We employ a powerful large vision-language model (LVLM), Qwen2.5-VL-72B \cite{qwenvl}, to detect whether a video contains sensitive information.
We sample frames at a rate of 1 FPS between the first and last frames of each video and input them into the LVLM to detect the presence of sensitive information. 
We filter out videos in which at least one frame contains such sensitive information, resulting in Ophora-28K, a privacy-preserved dataset comprising over 28K video-instruction pairs.
Then, we fine-tune the continual pre-trained model using Ophora-28K to enhance privacy while avoiding overwriting the previously learned spatial-temporal knowledge.

\section{Experiments}
%四个表格（1. 评分示意；2. 定量指标；3. 医生评价；4. 下游任务表格）
%2个图（1. 方法示意图；2. 生成视频示意图（不同的方法和消融的结果，5行））
% 训练的细节
%\subsection{Dataset and Implementation}

\noindent
\textbf{Dataset.}
Ophora-160K contains 162,185 video clip-instruction pairs extracted from 9,819 narrative videos of ophthalmic surgery. The average duration of all clips is 5.54 seconds. 
For data pre-processing, each clip was resized to a resolution of $720\times480$ and uniformly sampled to 49 frames. Clips with fewer than 49 frames were padded with all-zero frames until they reached 49 frames.
The dataset was split into 80\% and 20\% for training and testing, respectively. 
We constructed Ophora-28K, which contains 28,175 clip-instruction pairs from the training set.

\noindent
\textbf{Implementation.} 
For transfer pre-training, we employ AdamW optimizer with learning rate $1\times 10^{-4}$, batch size 128, and iteration number 65000 for sufficient knowledge transferring. 
For privacy-preserving fine-tuning, we employ the same optimizer with learning rate $5\times 10^{-5}$, batch size 128, and 4500 iterations to avoid knowledge overwriting. All experiments are conducted on A100 GPUs.

%\subsection{Evaluation of Synthesized Video Quality}

\noindent
\textbf{Quantitative Analysis.}
% 生成指标评估，相关setting + 结果 + 简要说明
We evaluated the quality of synthesized videos generated by our model, \textbf{Ophora}, against existing state-of-the-art surgical video generation models: \textbf{Endora} \cite{endora} and \textbf{Bora} \cite{bora} on the test set. 
We employ three metrics: Fr\'echet Inception Distance (FID) \cite{fid}, Fr\'echet Video Distance (FVD) \cite{fvd}, and CLIPScore (CS) \cite{clipscore} to evaluate the realism and video-text consistency of the synthesized videos.
The CS is calculated with the coefficient $\omega=100$ based on OphCLIP \cite{hu2024ophcliphierarchicalretrievalaugmentedlearning}, which aligns ophthalmic clips with narrative texts. 
%To evaluate the realism of the synthesized videos, we employ Fr\'echet Inception Distance (FID) and Fr\'echet Video Distance (FVD).
%To evaluate the consistency between the synthesized video and the input text prompt, we apply CLIPScore (CS) \cite{clipscore} based on OphCLIP \cite{hu2024ophcliphierarchicalretrievalaugmentedlearning}, which aligns ophthalmic clips with narrative texts. 
%We set the coefficient $\omega=100$ for the CLIPScore calculation.
% 可以考虑在这里引入一个temporal consistent的metric
%As Endora and Bora lack the capability to generate ophthalmic videos, 
We also fine-tune these models using Ophora-160K for comparison.

% 画实验表格，写实验结果，简单写了
As presented in Table \ref{tab:comparison}, Ophora achieved the best performance across all metrics compared to Endora and Bora. 
Ophora can generate ophthalmic videos with high visual fidelity, as measured by FID and FVD, due to inheriting rich spatial-temporal knowledge from the backbone pre-trained on large-scale natural videos, whereas the backbones of Endora and Bora lack such knowledge.
Moreover, Ophora can generate videos from text, but Endora lacks this capability, as it is an unconditional video generation model.
Meanwhile, the synthesized videos of Ophora achieve the highest correspondence with the input text, as they obtain the highest CS, even after these models are fine-tuned on our dataset. 
%如果引入temporal consistent的metric，可以再这里再加一句话分析其结果

% --感觉需要加一个dynamics相关的metric
\begin{table}[htbp]
    \caption{Comparison of synthesized video quality across different models based on quantitative metrics. \textbf{Bold} font denotes the best performance for each metric, and '-' indicates that CLIPScore (CS) was not calculated for this model.}
    \centering
    \begin{tabular}{ll|cccccccc}
        \hline \hline
        \multicolumn{2}{c|}{\multirow{2}{*}{Model}}  & \multicolumn{2}{c|}{Dataset setting} & \multicolumn{6}{c}{Metric} \\
        
        \cline{3-10}
        
        \multicolumn{2}{l|}{} & \multicolumn{1}{c}{OphVL \cite{hu2024ophcliphierarchicalretrievalaugmentedlearning}} & \multicolumn{1}{c|}{Ophora-160K} & \multicolumn{2}{c}{FID $\downarrow$} & \multicolumn{2}{c}{FVD $\downarrow$} & \multicolumn{2}{c}{CS $\uparrow$} \\

        \hline \hline
        % Endora results
        % 待测试
        \multicolumn{2}{l|}{Endora \cite{endora}} &  & \multicolumn{1}{c|}{} & \multicolumn{2}{c}{167.75} & \multicolumn{2}{c}{1433.29} & \multicolumn{2}{c}{-} \\ 

        \multicolumn{2}{l|}{Endora (w/ Ophora-160K)} &  & \multicolumn{1}{c|}{\checkmark}  & \multicolumn{2}{c}{60.50} & \multicolumn{2}{c}{990.30} & \multicolumn{2}{c}{-} \\ 

        \hline

        % Bora results 
        % 待测试
        \multicolumn{2}{l|}{Bora \cite{bora}} &  &  \multicolumn{1}{c|}{} & \multicolumn{2}{c}{138.30} & \multicolumn{2}{c}{1761.42} & \multicolumn{2}{c}{12.68} \\ 

        \multicolumn{2}{l|}{Bora (w/ Ophora-160K)} &  & \multicolumn{1}{c|}{\checkmark}  & \multicolumn{2}{c}{49.74} & \multicolumn{2}{c}{604.20} & \multicolumn{2}{c}{32.02} \\ 

        \hline

        %Ophora results
        % 待推理
       \multicolumn{2}{l|}{CogVideoX-2b \cite{yang2024cogvideox}} &  & \multicolumn{1}{c|}{}  & \multicolumn{2}{c}{138.20} & \multicolumn{2}{c}{871.16} & \multicolumn{2}{c}{5.87} \\ 

        %训练中
       \multicolumn{2}{l|}{CogVideoX-2b (w/ OphVL)} & \checkmark & \multicolumn{1}{c|}{}  & \multicolumn{2}{c}{61.48} & \multicolumn{2}{c}{532.47} & \multicolumn{2}{c}{33.65} \\ 

        %推理中
       \multicolumn{2}{l|}{Ophora (TPT-only)} & & \multicolumn{1}{c|}{\checkmark} & \multicolumn{2}{c}{42.16} & \multicolumn{2}{c}{441.09} & \multicolumn{2}{c}{37.03} \\ 
       
        \multicolumn{2}{l|}{Ophora } &  & \multicolumn{1}{c|}{\checkmark}  & \multicolumn{2}{c}{\textbf{33.72}} & \multicolumn{2}{c}{\textbf{276.96}} & \multicolumn{2}{c}{\textbf{39.19}} \\

        \hline \hline
    \end{tabular}
    
    \label{tab:comparison}
\end{table}
% 可以给Bora和Endora加个星号，说明是基于我们的数据微调的
\begin{figure}[!t]
    \includegraphics[width=\textwidth]{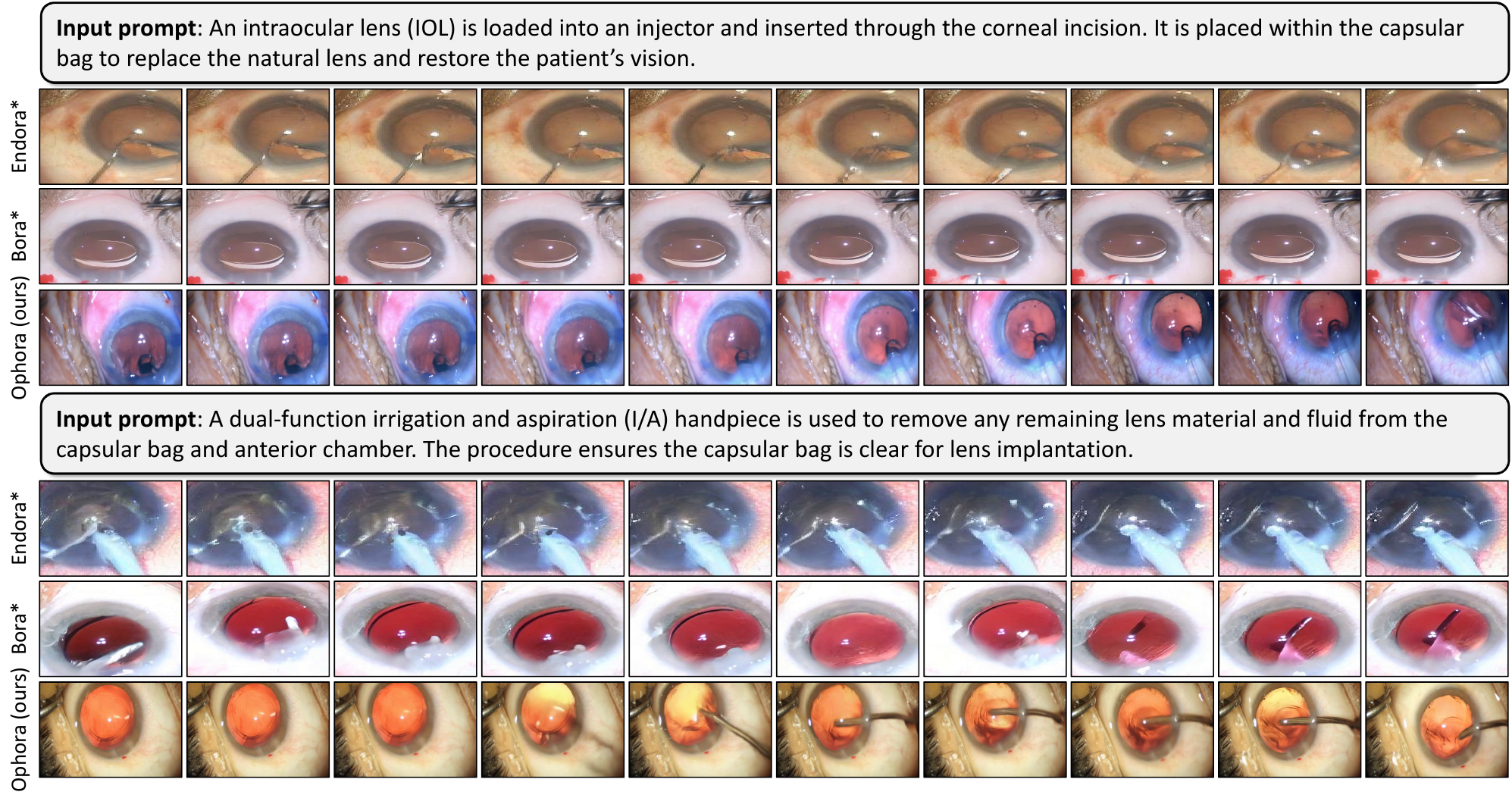}
    \caption{Synthesized video frames from the input text prompts of different models. ‘*’ denotes that this model was fine-tuned on the proposed Ophora-160K.}
    \label{fig:visualize}
\end{figure}

%展示结果，并简单描述一下生成的结果的质量
\noindent
\textbf{Visualization.}
We also present the synthesized video frames generated from the input text prompts in Fig. \ref{fig:visualize}. 
As Endora cannot generate videos from text, we randomly select some synthesized videos for visualization.
The synthesized videos from Ophora demonstrate realistic and detailed surgical scenes with proper instruments and coherent surgical actions that follow the input prompts.
Although other models are fine-tuned on our dataset, they exhibit limited instruction-following capabilities, as evidenced by the inconsistency between the instruments and actions described in the prompt and those in the synthesized videos.

\begin{figure}[!t]
    \includegraphics[width=\textwidth]{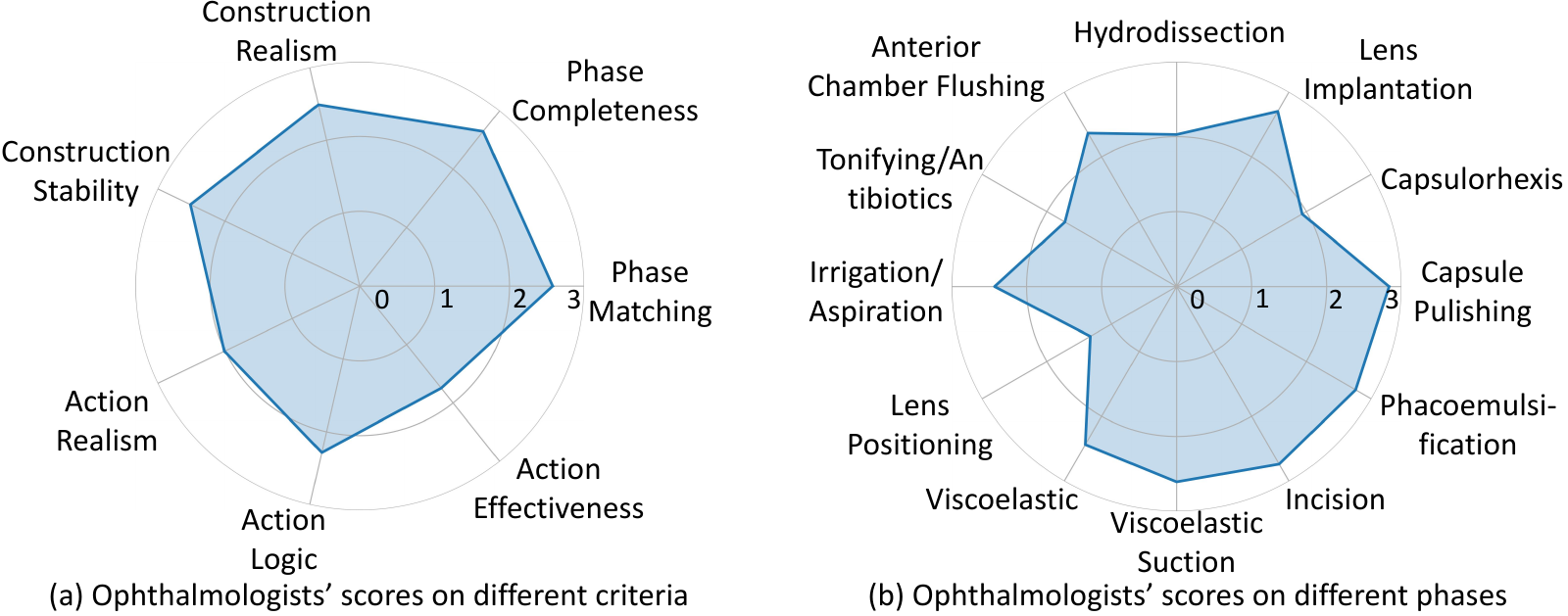}
    \caption{Ophthalmologists' scores on different criteria (a) and surgical phases (b).}
    \label{fig:radfig}
\end{figure}

\noindent
\textbf{Ophthalmologist Feedback.}
% 医生评估，相关setting + 结果 + 简要说明
We generated 600 videos based on the surgical phase labels of Cataract-1K dataset \cite{catar1k} with 50 videos for each label.
These videos were synthesized based on generation instructions that were written by three ophthalmologists according to the phase labels.
Then, three ophthalmologists evaluated the authenticity of these synthesized videos based on seven criteria, including: 
1) \textbf{Phase Matching} that reflects the consistency between the generated video and the corresponding phase label; 
2) \textbf{Phase Completeness}, which indicates whether the video presents a complete phase;
3) \textbf{Construction Realism} representing the authenticity of the anatomical structures or instruments; 
4) \textbf{Construction Stability} indicating whether the anatomical structures or instruments undergo unrealistic deformations;
5) \textbf{Action Realism}, which represents the consistency between the surgical actions in videos and actual procedures;
6) \textbf{Action Logic} reflecting the consistency of the sequential order of surgical actions with actual orders;
7) \textbf{Action Effectiveness,} meaning the realism of the deformation effects on the corresponding tissues caused by surgical actions.
Each criterion is scored on a scale of 0 to 3, representing different levels of realism, where a score of 0 indicates 0\% $\sim$ 10\% realism, 1 corresponds to 10\%$\sim$50\%, 2 to 50\%$\sim$90\%, and 3 to 90\%$\sim$100\% realism, respectively.

The results are shown in Fig. \ref{fig:radfig}. 
For most phases, Ophora can generate videos being consistent with the phases given phase-related generation instructions, with no erroneous deformation of anatomical structures or instruments and realistic actions, as evidenced by the high scores on different criteria and phases.
%The synthesized videos demonstrate realistic surgical actions that closely follow the actual sequence of operational logic, as reflected by the scores on AR, AL and AE.

\begin{table}[!htbp]
  \centering
  \small
  \caption{Comparison of the top-1 and top-5 accuracy of two classifiers on the validation and test sets of OphNet, including phase and operation-based classification tasks, under three training data configurations. \textbf{Bold} denotes the best performance for each split.}
  \resizebox{\linewidth}{!}{
    \begin{tabular}{l|l|cc|cc|cc|cc}  % 修改列对齐方式，确保“Top-1”列居中对齐
    \hline \hline
    \multirow{3}{*}{Training data}      & \multirow{3}{*}{Classifier}       & \multicolumn{4}{c|}{Val}      & \multicolumn{4}{c}{Test} \\
    \cline{3-10}
          &       & \multicolumn{2}{c|}{Phase} & \multicolumn{2}{c|}{Operation} & \multicolumn{2}{c|}{Phase} & \multicolumn{2}{c}{Operation} \\
    \cline{3-10}
          &       & \multicolumn{1}{r}{Top-1} & \multicolumn{1}{c|}{Top-5} & \multicolumn{1}{r}{Top-1} & \multicolumn{1}{c|}{Top-5} & \multicolumn{1}{r}{Top-1} & \multicolumn{1}{c|}{Top-5} & \multicolumn{1}{r}{Top-1} & \multicolumn{1}{c}{Top-5} \\
    \hline \hline
    \multirow{2}{*}{Source} 
          & SlowFast~\cite{feichtenhofer2019slowfast} & 34.55  & 70.24  & 26.93  & 65.11  & 37.04  & 72.69  & 27.21  & 67.26  \\
          %& SlowFast~\cite{feichtenhofer2019slowfast} & \multicolumn{1}{r}{34.55}  & \multicolumn{1}{r}{34.55}  & \multicolumn{1}{r}{34.55}  & \multicolumn{1}{r}{34.55}  & \multicolumn{1}{r}{34.55}  & \multicolumn{1}{r}{34.55}  & \multicolumn{1}{r}{34.55} & \multicolumn{1}{r}{34.55}  \\
          & MViTv2~\cite{li2022mvitv2} & 36.24  & 72.54  & 27.89  & 65.86  & 37.92  & 74.31  & 28.56  & 68.32  \\
    \hline
    \multicolumn{1}{l|}{\multirow{2}{*}{Source $+$ Bora}} 
          & SlowFast & 36.35  & 72.88  & 28.58  & 67.52  & 39.24  & 74.94  & 28.88  & 69.53  \\
          & MViTv2 & 37.43  & 73.62  & 29.59  & 68.34  & 39.26  & 75.76  & 30.44  & 70.32  \\
    \hline
    \multicolumn{1}{l|}{\multirow{2}{*}{Source $+$ Ophora}} 
          & SlowFast & 38.55  & 73.23  & 30.81  & 69.59  & 41.05  & 77.43  & 31.10  & 72.01  \\
          & MViTv2 & \textbf{40.15}  & \textbf{76.52}  & \textbf{32.80}  & \textbf{70.28}  & \textbf{42.24}  & \textbf{78.56}  & \textbf{33.62}  & \textbf{73.27}  \\
    \hline \hline
    \end{tabular}%
    }
  \label{tab:da}%
\end{table}

%\subsection{Data Augmentation on Downstream Task}
\noindent
\textbf{Data Augmentation on Downstream Task.}
We evaluate the impact of synthesized videos from Ophora as augmented data on OphNet \cite{hu2024ophnetlargescalevideobenchmark}, a large-scale benchmark for ophthalmic surgical workflow understanding, featuring 14,674 instances annotated with 52 surgical phases and 17,508 with 106 operations. 
%Due to its fine-grained annotations, OphNet exhibits a severe long-tail distribution, which greatly affects classification performance.
%We generated 100 videos per category—excluding the “Non-functional Segment” and “Step Interval” classes—for use as augmentation data.
% 如果空间不足可以去掉I3D来换取空间
We generated 100 videos for each category and compared two clip-level surgical workflow recognition models: SlowFast \cite{feichtenhofer2019slowfast} and MViTv2 \cite{li2022mvitv2}, under three training data configurations: 1) \textbf{source}, using only real videos from the training set of OphNet; 2) \textbf{source $+$ Bora}, including real videos and synthesized videos from Bora; and 3) \textbf{source $+$ Ophora}, including real videos and synthesized videos from Ophora. 
%We downsample all videos to 5 FPS. 
We employed Adam optimizer with learning rate 0.1, epoch number 35, batch size 64 for training SlowFast, and batch size 32 for training MViTv2. We uniformly sampled 8 frames for each video during training.
%All models are trained for 30 epochs, with 5 warm-up epochs. 
%Each video clip is sampled with 8 frames at a sampling interval of 5.

We report the Top-1 and Top-5 accuracy of two classifiers on the validation and test sets across different training setups for phase-based and operation-based classification tasks in Table \ref{tab:da}. 
Although synthesized videos from Bora yield moderate accuracy gains compared to the source setting, leveraging the videos from Ophora achieves the best performance across all classifiers on both classification tasks. 
Notably, MViTv2 achieves the highest boost in phase-level Top-1 accuracy on the test set (37.92\% → 42.24\%) due to the adoption of synthesized videos from Ophora. 
The results indicate that Ophora can generate diverse and effective videos for developing an AI model for ophthalmic workflow understanding.

% In Tab.~\ref{tab:enhen}, we compare the performance of three video classification models (I3D, Slowfast, MViTv2) on the original dataset (Source) versus two generated data augmentations (Bora and Ophora). While Bora augmentation provides moderate improvements over the Source baseline, Ophora achieves more substantial gains across both phase- and operation-level tasks. Notably, MViTv2 benefits the most with a 4.32\% absolute increase in phase-level Top-1 accuracy on the test set (from 37.92\% to 42.24\%), clearly outperforming Bora. These results confirm that Ophora generates more diverse and effective training samples, thereby enhancing model generalization beyond both the baseline and Bora-augmented settings.

%补充结果分析
%加一句话提一下workflow 理解的意义
% We validated the downstream performance of Ophora on the OphNet [xx], a large-scale benchmark for ophthalmic surgical workflow understanding. OphNet collects over 2K ophthalmic videos with annotations for 150 unique surgical phases and operations. 
% We generated 100 videos for each category as augmentation data and trained state-of-the-art phase recognition models xxx [xx] using three training data settings for comparison: 1) real videos of training set (baseline); 2) baseline + Bora synthesized videos; 3) baseline + Ophora synthesized videos.
% 这里可以再简单提一下训练分类器时的learning rate, batch size, optimizer, epochs

% We reported the accuracy and F1-score of three training settings on the test in Table xx.
%接下来写实验结果分析

\noindent
\textbf{Ablation Study.}
We compared the performance of Ophora with: 
1) the original CogVideoX-2b; 
2) CogVideoX-2b (w/ OphVL) representing directly fine-tuning CogVideoX-2b on OphVL without applying our data curation;
and 3) Ophora (TPT-only), where we only conducted transfer pre-training (TPT), as shown in Table \ref{tab:comparison}.
Although fine-tuning CogVideoX-2b on OphVL enables ophthalmic video generation, the performance is limited compared to Ophora due to redundant information in narratives and low temporal dynamics quality of videos.
In contrast, performing TPT on the same backbone using Ophora-160K further improves performance. 
Surprisingly, after conducting privacy-preserving fine-tuning, Ophora achieves the best performance, as videos without sensitive information may correspond better with the narrative text.

\section{Conclusion}
This paper presents Ophora, a pioneering model that can generate ophthalmic surgical videos following input generation instructions. 
To achieve this, we propose a comprehensive data curation pipeline to convert narrative videos into a large-scale, high-quality video-instruction dataset, Ophora-160K. 
Then, we propose a progressive video-instruction tuning approach to transfer spatial-temporal knowledge from a T2V model pre-trained on natural video-text datasets for privacy-preserved ophthalmic video generation.
Ophora can generate realistic and reliable ophthalmic videos based on user instructions, demonstrating significant potential for developing a general surgical AI system.
In the future, we will explore generating videos of more surgery types with longer durations.

%
% ---- Bibliography ----
%
\bibliographystyle{splncs04}
\bibliography{egbib}

\begin{thebibliography}{10}
\providecommand{\url}[1]{\texttt{#1}}
\providecommand{\urlprefix}{URL }
\providecommand{\doi}[1]{https://doi.org/#1}

\bibitem{qwenvl}
Bai, S., Chen, K., Liu, X., Wang, J., Ge, W., Song, S., Dang, K., Wang, P., Wang, S., Tang, J., et~al.: Qwen2. 5-vl technical report. arXiv preprint arXiv:2502.13923  (2025)

\bibitem{cepolina2024review}
Cepolina, F., Razzoli, R.: Review of robotic surgery platforms and end effectors. Journal of Robotic Surgery  \textbf{18}(1), ~74 (2024)

\bibitem{cheng2023deep}
Cheng, Y., Liu, L., Wang, S., Jin, Y., Sch{\"o}nlieb, C.B., Aviles-Rivero, A.I.: Why deep surgical models fail?: Revisiting surgical action triplet recognition through the lens of robustness. In: International Workshop on Trustworthy Machine Learning for Healthcare. pp. 177--189. Springer (2023)

\bibitem{cho2024surgentextguideddiffusionmodel}
Cho, J., Schmidgall, S., Zakka, C., Mathur, M., Kaur, D., Shad, R., Hiesinger, W.: Surgen: Text-guided diffusion model for surgical video generation. arXiv preprint arXiv:2408.14028  (2024)

\bibitem{feichtenhofer2019slowfast}
Feichtenhofer, C., Fan, H., Malik, J., He, K.: Slowfast networks for video recognition. In: Proceedings of the IEEE/CVF international conference on computer vision. pp. 6202--6211 (2019)

\bibitem{catar1k}
Ghamsarian, N., El-Shabrawi, Y., Nasirihaghighi, S., Putzgruber-Adamitsch, D., Zinkernagel, M., Wolf, S., Schoeffmann, K., Sznitman, R.: Cataract-1k dataset for deep-learning-assisted analysis of cataract surgery videos. Scientific Data  \textbf{11}(1), ~373 (2024)

\bibitem{he2021robotoph}
He, B., de~Smet, M.D., Sodhi, M., Etminan, M., Maberley, D.: A review of robotic surgical training: establishing a curriculum and credentialing process in ophthalmology. Eye  \textbf{35}(12),  3192--3201 (2021)

\bibitem{clipscore}
Hessel, J., Holtzman, A., Forbes, M., Bras, R.L., Choi, Y.: Clipscore: A reference-free evaluation metric for image captioning. arXiv preprint arXiv:2104.08718  (2021)

\bibitem{fid}
Heusel, M., Ramsauer, H., Unterthiner, T., Nessler, B., Hochreiter, S.: Gans trained by a two time-scale update rule converge to a local nash equilibrium. In: Guyon, I., Luxburg, U.V., Bengio, S., Wallach, H., Fergus, R., Vishwanathan, S., Garnett, R. (eds.) Advances in Neural Information Processing Systems. vol.~30. Curran Associates, Inc. (2017)

\bibitem{ddpm}
Ho, J., Jain, A., Abbeel, P.: Denoising diffusion probabilistic models. In: Larochelle, H., Ranzato, M., Hadsell, R., Balcan, M., Lin, H. (eds.) Advances in Neural Information Processing Systems. vol.~33, pp. 6840--6851. Curran Associates, Inc. (2020)

\bibitem{hu2024ophnetlargescalevideobenchmark}
Hu, M., Xia, P., Wang, L., Yan, S., Tang, F., Xu, Z., Luo, Y., Song, K., Leitner, J., Cheng, X., Cheng, J., Liu, C., Zhou, K., Ge, Z.: Ophnet: A large-scale video benchmark for ophthalmic surgical workflow understanding. In: Leonardis, A., Ricci, E., Roth, S., Russakovsky, O., Sattler, T., Varol, G. (eds.) Computer Vision -- ECCV 2024. pp. 481--500. Springer Nature Switzerland, Cham (2025)

\bibitem{hu2024ophcliphierarchicalretrievalaugmentedlearning}
Hu, M., Yuan, K., Shen, Y., Tang, F., Xu, X., Zhou, L., Li, W., Chen, Y., Xu, Z., Peng, Z., et~al.: Ophclip: Hierarchical retrieval-augmented learning for ophthalmic surgical video-language pretraining. arXiv preprint arXiv:2411.15421  (2024)

\bibitem{dynamicsbench}
Huang, Z., He, Y., Yu, J., Zhang, F., Si, C., Jiang, Y., Zhang, Y., Wu, T., Jin, Q., Chanpaisit, N., Wang, Y., Chen, X., Wang, L., Lin, D., Qiao, Y., Liu, Z.: Vbench: Comprehensive benchmark suite for video generative models. In: 2024 IEEE/CVF Conference on Computer Vision and Pattern Recognition (CVPR). pp. 21807--21818 (2024)

\bibitem{Laparoscopicgen}
Iliash, I., Allmendinger, S., Meissen, F., K{\"u}hl, N., R{\"u}ckert, D.: Interactive generation of laparoscopic videos with diffusion models. In: Mukhopadhyay, A., Oksuz, I., Engelhardt, S., Mehrof, D., Yuan, Y. (eds.) Deep Generative Models. pp. 109--118. Springer Nature Switzerland, Cham (2025)

\bibitem{laborconsump}
Jin, Y., Dou, Q., Chen, H., Yu, L., Qin, J., Fu, C.W., Heng, P.A.: Sv-rcnet: Workflow recognition from surgical videos using recurrent convolutional network. IEEE Transactions on Medical Imaging  \textbf{37}(5),  1114--1126 (2018)

\bibitem{miradata}
Ju, X., Gao, Y., Zhang, Z., Yuan, Z., Wang, X., ZENG, A., Xiong, Y., Xu, Q., Shan, Y.: Miradata: A large-scale video dataset with long durations and structured captions. In: Globerson, A., Mackey, L., Belgrave, D., Fan, A., Paquet, U., Tomczak, J., Zhang, C. (eds.) Advances in Neural Information Processing Systems. vol.~37, pp. 48955--48970. Curran Associates, Inc. (2024)

\bibitem{endora}
Li, C., Liu, H., Liu, Y., Feng, B.Y., Li, W., Liu, X., Chen, Z., Shao, J., Yuan, Y.: Endora: Video generation models as endoscopy simulators. In: Linguraru, M.G., Dou, Q., Feragen, A., Giannarou, S., Glocker, B., Lekadir, K., Schnabel, J.A. (eds.) Medical Image Computing and Computer Assisted Intervention -- MICCAI 2024. pp. 230--240. Springer Nature Switzerland, Cham (2024)

\bibitem{li2022mvitv2}
Li, Y., Wu, C.Y., Fan, H., Mangalam, K., Xiong, B., Malik, J., Feichtenhofer, C.: Mvitv2: Improved multiscale vision transformers for classification and detection. In: Proceedings of the IEEE/CVF conference on computer vision and pattern recognition. pp. 4804--4814 (2022)

\bibitem{t2voriginal}
Li, Y., Min, M., Shen, D., Carlson, D., Carin, L.: Video generation from text. Proceedings of the AAAI Conference on Artificial Intelligence  \textbf{32}(1) (Apr 2018)

\bibitem{price2019privacy}
Price, W.N., Cohen, I.G.: Privacy in the age of medical big data. Nature medicine  \textbf{25}(1),  37--43 (2019)

\bibitem{t5encoder}
Raffel, C., Shazeer, N., Roberts, A., Lee, K., Narang, S., Matena, M., Zhou, Y., Li, W., Liu, P.J.: Exploring the limits of transfer learning with a unified text-to-text transformer. J. Mach. Learn. Res.  \textbf{21}(1) (Jan 2020)

\bibitem{bora}
Sun, W., You, X., Zheng, R., Yuan, Z., Li, X., He, L., Li, Q., Sun, L.: Bora: Biomedical generalist video generation model. arXiv preprint arXiv:2407.08944  (2024)

\bibitem{THIA2019570videorecord}
Thia, B.C., Wong, N.J., Sheth, S.J.: Video recording in ophthalmic surgery. Survey of Ophthalmology  \textbf{64}(4),  570--578 (2019)

\bibitem{fvd}
Unterthiner, T., Van~Steenkiste, S., Kurach, K., Marinier, R., Michalski, M., Gelly, S.: Towards accurate generative models of video: A new metric \& challenges. arXiv preprint arXiv:1812.01717  (2018)

\bibitem{aiinsurg}
Varghese, C., Harrison, E.M., O’Grady, G., Topol, E.J.: Artificial intelligence in surgery. Nature Medicine  \textbf{30}(5),  1257--1268 (2024)

\bibitem{qwen25}
Yang, A., Yang, B., Zhang, B., Hui, B., Zheng, B., Yu, B., Li, C., Liu, D., Huang, F., Wei, H., et~al.: Qwen2. 5 technical report. arXiv preprint arXiv:2412.15115  (2024)

\bibitem{yang2024cogvideox}
Yang, Z., Teng, J., Zheng, W., Ding, M., Huang, S., Xu, J., Yang, Y., Hong, W., Zhang, X., Feng, G., et~al.: Cogvideox: Text-to-video diffusion models with an expert transformer. arXiv preprint arXiv:2408.06072  (2024)

\bibitem{hecvl}
Yuan, K., Srivastav, V., Navab, N., Padoy, N.: Hecvl: Hierarchical video-language pretraining for zero-shot surgical phase recognition. In: Linguraru, M.G., Dou, Q., Feragen, A., Giannarou, S., Glocker, B., Lekadir, K., Schnabel, J.A. (eds.) Medical Image Computing and Computer Assisted Intervention -- MICCAI 2024. pp. 306--316. Springer Nature Switzerland, Cham (2024)

\bibitem{transferlearning}
Zhuang, F., Qi, Z., Duan, K., Xi, D., Zhu, Y., Zhu, H., Xiong, H., He, Q.: A comprehensive survey on transfer learning. Proceedings of the IEEE  \textbf{109}(1),  43--76 (2021)

\end{thebibliography}

\end{document}